\documentclass[sigconf,authorversion]{acmart}
\AtBeginDocument{%
  }

\setcopyright{acmlicensed}
\copyrightyear{2026}
\acmYear{2026}
\acmDOI{XXXXXXX.XXXXXXX}
\acmConference[VRST '26]{Symposium on Virtual Reality Software and Technology}{November 16--18,
  2026}{Sendai, Japan}
\acmISBN{978-1-4503-XXXX-X/2018/06}

\usepackage{pgfplots}
\pgfplotsset{compat=1.18}
\usepackage{subcaption}
\usepackage{graphicx}

\begin{document}

\title[Listening and Mirroring: The Effects of Verbal Attunement and Behavioral Mimicry on Social Perceptions of Embodied AI Agents]{Listening and Mirroring: The Effects of Verbal Attunement and Behavioral Mimicry on Social and Empathic Perceptions of Embodied AI Agents in VR}

\author{Nathalia Gomez}
\email{nmg88@drexel.edu}
\orcid{0009-0005-1693-9535}
\affiliation{%
  \institution{Drexel University}
  \city{Philadelphia}
  \state{PA}
  \country{USA}
}

\author{Haig Shamlian}
\email{hks55@drexel.edu}
\affiliation{%
  \institution{Drexel University}
  \city{Philadelphia}
  \state{PA}
  \country{USA}
}

\author{Omar Khan}
\email{oak44@drexel.edu}
\affiliation{%
  \institution{Drexel University}
  \city{Philadelphia}
  \state{PA}
  \country{USA}
}

\author{Tiffany D. Do}
\email{td848@drexel.edu}
\affiliation{%
 \institution{Drexel University}
 \city{Philadelphia}
 \state{PA}
 \country{USA}
 }

\renewcommand{\shortauthors}{Gomez et al.}

\begin{abstract}
As embodied agents take on increasingly social and relational roles in VR, visual realism and embodiment alone may be insufficient; users must also perceive these agents as emotionally attuned, supportive, and humanlike. Prior work suggests that verbal attunement and nonverbal mimicry can each improve users' social evaluations of embodied agents. However, behavioral mimicry has largely been studied outside of real-time, conversational AI interactions, leaving limited understanding of how users respond when an agent simultaneously generates contextually responsive dialogue and adapts its nonverbal behavior during an immersive conversation. To address this gap, we developed an embodied AI counselor that combines conversational AI with real-time facial-expression and posture mimicry, while producing either verbally attuned or neutral responses. We evaluated the system in a $2 \times 2$ within-subjects study with 20 participants, manipulating verbal attunement and behavioral mimicry. Results showed that verbal attunement was the most reliable driver of perceived empathy. Behavioral mimicry showed a marginal relationship with perceived humanness, while greater mimicry exposure showed preliminary, exploratory positive associations with empathy, positivity, and humanness, particularly among female participants. Together, these findings show that multimodal synchrony is not a simple additive strategy for designing empathic conversational agents in VR and underscore the need to consider how verbal and nonverbal behaviors are combined during real-time interaction.
\end{abstract}



\begin{teaserfigure}
  \includegraphics[width=\textwidth]{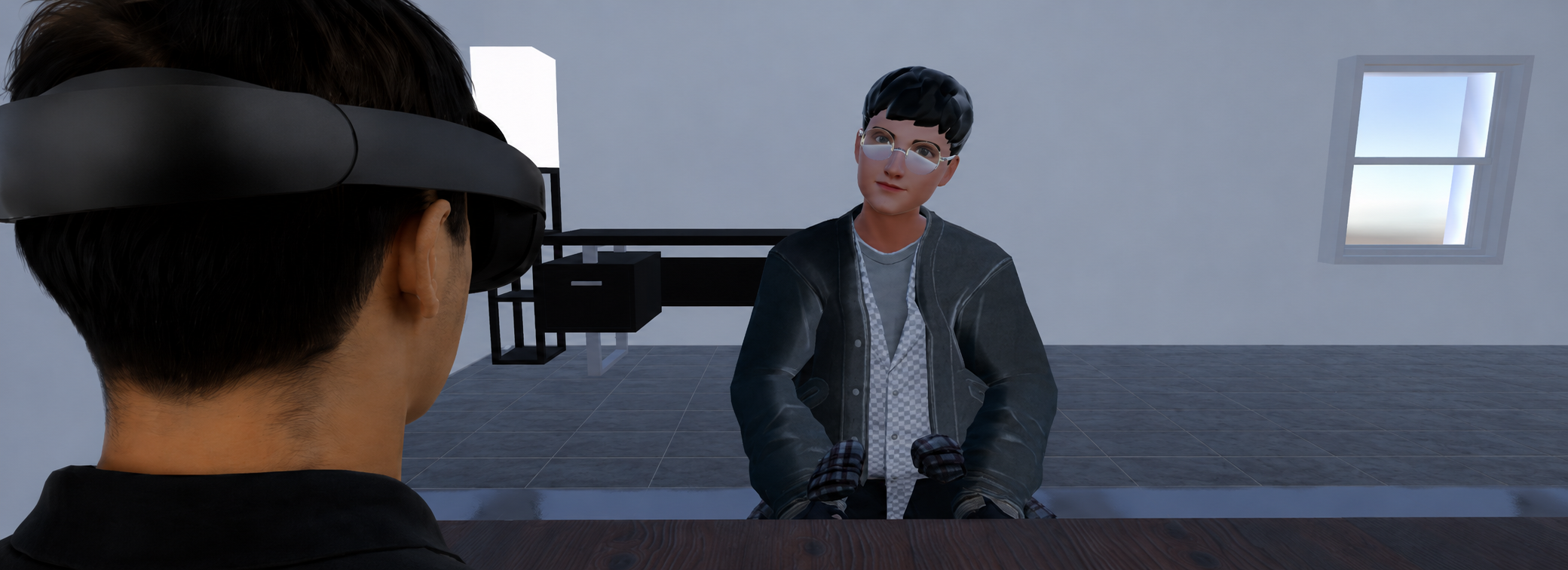}
  \caption{Representation of a participant speaking to the embodied AI counselor in VR}
  \Description{Enjoying the baseball game from the third-base
  seats. Ichiro Suzuki preparing to bat.}
  \label{fig:teaser}
\end{teaserfigure}

\maketitle

\section{Introduction}

Embodied AI agents are becoming an increasingly common interface for social interaction in XR~\cite{Yang2025}. Rather than functioning only as tools or assistants, these agents are beginning to appear as tutors, collaborators, coaches, healthcare aides, and companions~\cite{kruse_would_2023,rings_ivas_2025,yu_embodied_2026,kim_reducing_2020}. In these roles, it is not enough for an agent to respond correctly. Users may also need to perceive the agent as attentive, emotionally responsive, and socially present. Prior work suggests that virtual agents have the potential to support long-term human--AI relationships that resemble social affiliation~\cite{Fung2025,Paiva2017}.  This is especially important in immersive VR, where agents occupy the same spatial environment as the user and can communicate through both speech and embodied behavior.

Recent advances in large language models have made real-time conversation with AI agents more natural and flexible. At the same time, VR allows these agents to communicate through nonverbal cues such as posture, facial expression, gaze, and movement. Together, these developments create new opportunities for designing AI agents that feel more empathic and humanlike. However, they also raise an important design question: when an AI agent is both conversational and embodied, which cues actually shape users' perceptions of empathy?

Prior work suggests that verbal attunement can increase perceived empathy, particularly when responses include validation, personalized active listening, nonjudgmental language, and emotional reflection ~\cite{Ruben2026}. Other work on behavioral mimicry suggests that agents who mirror users' nonverbal behavior can be evaluated more positively and may support feelings of rapport, affiliation, or social presence~\cite{Bailenson2005,Hale2016}. However, the behavioral mimicry literature largely predates conversational AI systems capable of sustaining contextually responsive dialogue. These earlier studies established the social effects of nonverbal matching, but did not examine how users interpret mimicry when an agent is simultaneously producing responsive, context-sensitive language. It therefore remains unclear how verbal attunement and behavioral mimicry jointly shape users' perceptions during natural, real-time conversations with embodied AI agents in VR.
To investigate this gap, we developed an embodied AI counselor in VR that could produce either verbally attuned or neutral responses and could either mimic or not mimic selected participant behaviors, including facial expression, head tilt, and posture. We evaluated the system in a 2 x 2 within-subjects study with 20 participants, manipulating verbal attunement and behavioral mimicry. We measured participants' perceptions of empathy, general positivity, humanness, eeriness, and verbal attunement, and also collected open-ended feedback after each interaction.

This work is guided by the following research questions:
\begin{itemize}
\item \textbf{RQ1:} How does verbal attunement affect users' perceptions of an embodied AI agent's empathy, positivity, humanness, and eeriness in VR?
\item \textbf{RQ2:} How does behavioral mimicry affect users' perceptions of an embodied AI agent's empathy, positivity, humanness, and eeriness in VR?
\item \textbf{RQ3:} How do verbal attunement and behavioral mimicry operate together in shaping users' perceptions of an embodied AI agent in VR?
\end{itemize}

Our findings show that verbal attunement was the clearest driver of perceived empathy, while behavioral mimicry played a different role by contributing more to perceived humanness and exposure-based patterns. These results suggest that multimodal synchrony should not be treated as a simple additive strategy for making VR agents empathic. Instead, verbal attunement, embodied behavior, and conversational timing appear to shape different parts of the user's social experience with an AI agent.

\section{Related Work}

\subsection{Embodied AI in VR for Healthcare and Collaboration}
Embodied AI can be categorized into graphical representations (accessible via screens and VR/AR interfaces), and physical representations such as robots~\cite{liu_screens_2025, mir_embodied_2026}.
Graphical representations of embodied AI (e.g., virtual humans) have long been a topic of interest in the VR/AR research community~\cite{kim_revisiting_2018, weidner_systematic_2023}, and have been shown to lead to positive outcomes in healthcare, such as mental health treatment~\cite{kruse_would_2023}, and telecare~\cite{winkler_avatar_2025}.
Rings et al.~\cite{rings_ivas_2025} compared graphically embodied virtual agents for mental health in a mobile app versus in VR, and found tradeoffs in the user experience of each. 
Yu et al.~\cite{yu_embodied_2026} showed that embodied agents can enhance efficiency in aphasia rehabilitation. 
Beyond healthcare, embodied AI in VR has shown promise in other contexts, such as enhancing human-AI collaboration and teaming. For instance, Kim et al.~\cite{kim_reducing_2020} found that, by providing non-verbal social cues, embodied virtual assistants enhanced performance and reduced task load in collaborative decision making, compared to disembodied assistants. Embodied AI systems have also shown value in fostering human-human collaboration, leading to positive outcomes in domains such as education~\cite{earle-randell_how_2025}.
Embodied AI present various dimensions which have shown to be instrumental in shaping user experience, such as voice~\cite{do_effects_2022}, appearance~\cite{volante_effects_2016}, gestures~\cite{kim_dominance_2026}, and personality~\cite{sonlu_effects_2025}. 

This work shows that embodied AI is increasingly being used in applied VR contexts where agents are expected to interact with users socially. This makes the design of embodied agents consequential because their language, movement, and appearance can shape how users interpret them as social partners. In this paper, we focus on two such cues, verbal attunement and behavioral mimicry, and examine how they affect users' perceptions of an embodied AI counselor in VR.

\subsection{Importance of Empathy in Human-AI Interaction}
In a recent systematic review of embodied conversational agents (ECAs) in XR, Yang et al. found that training and education were the most common application areas, followed by therapy and counseling~\cite{Yang2025}. This suggests that ECAs are increasingly being designed for roles that involve guidance, support, and social interaction rather than only task execution. As these agents appear as teachers, coaches, counselors, and companions, it is not enough for them to respond correctly. Users may also need to perceive the agent as attentive, emotionally responsive, and socially present. This is especially important in counseling contexts, where therapist empathy is a reliable predictor of therapeutic outcome and client experience~\cite{Elliott2018}. Moreover, prior work on relational agents has shown that social-emotional behaviors, including empathy, trust-building, and relationship maintenance strategies, are important for supporting long-term human--computer relationships~\cite{Bickmore2005}.

In immersive VR, conversational agents have an additional layer of social expression that text-based systems do not: embodiment. Through facial expression, posture, movement, gaze, and spatial presence, VR agents can communicate not only through what they say, but also through how they appear to respond in the shared environment. Fung et al. argue that embodied AI agents can enable more natural and intuitive interaction by grounding communication in shared physical and social context, and may evoke forms of social and emotional engagement that disembodied systems cannot support in the same way~\cite{Fung2025}. Similarly, Paiva et al.'s survey of empathic virtual agents and robots shows that empathic agents are often perceived as more likable, trustworthy, and caring, and that some can support long-term socioemotional relationships with human partners~\cite{Paiva2017}.

Together, this work points to the promise of empathic embodied agents, but it also raises a more specific design question: which cues make an embodied AI agent feel empathic? In this paper, we focus on two forms of attunement that are especially relevant to real-time VR conversation. The first is behavioral mimicry, in which an embodied agent mirrors aspects of the user's nonverbal behavior, such as facial expression, head movement, or posture. The second is verbal attunement, in which the agent's language reflects emotional understanding through validation, acknowledgment, and context-specific responses. Prior work suggests that both verbal and nonverbal cues can shape social evaluations of agents, but they are often studied separately. Less is known about how behavioral mimicry and verbal attunement operate together in a real-time conversation with an embodied AI agent. This paper addresses that gap by examining how these two forms of attunement shape perceived empathy, positivity, humanness, and eeriness in an embodied AI counselor.

\subsection{Behavioral Mimicry in Embodied Agents}

One nonverbal cue often associated with social connection is behavioral mimicry, or the mirroring of another person's facial expressions, posture, gestures, or movement. In human-human interaction, mimicry has been linked to affiliation, rapport, and smoother social interaction, often described through the ``chameleon effect''~\cite{Lakin2003}. This idea has also been extended to artificial agents. Bailenson and Yee~\cite{Bailenson2005} showed that an embodied agent in immersive VR that subtly mimicked a user's head movements was rated more positively and was more persuasive than a non-mimicking agent.

More recent work has examined how embodied systems can use mimicry to shape social perception. De Borst and de Gelder~\cite{deBorst2015} review evidence that emotional expressions by humanlike avatars can produce responses similar to those elicited by humans, including facial mimicry in observers. Extending mimicry beyond humanlike agents, Nam et al.~\cite{Nam2025} found that an AR virtual dog that mimicked users' exercise movements improved user experience, engagement, social presence, and social attachment compared to random-behavior and no-dog conditions.

Together, these studies suggest that behavioral mimicry can shape social evaluations of embodied agents, including persuasion, rapport, attachment, and social presence. However, it remains unclear whether these effects extend to perceived empathy in emotionally reflective interaction or primarily support broader impressions of social connection and responsiveness. Prior studies also focus on short interactions, scripted tasks, or mimicry separated from natural spoken conversation. In this study, we examine behavioral mimicry during real-time, emotionally reflective VR conversation by enabling the counselor to mirror selected participant facial expressions, head tilts, and posture changes.

\subsection{LLMs and Verbal Attunement}
Recent work suggests that LLMs and chatbots can generate responses that users perceive as empathic, especially in text-based or health-related contexts. Ayers et al.~\cite{Ayers2023} found that ChatGPT responses to patient questions were rated higher in both quality and empathy than physician responses. Similarly, Liu and Sundar found that a health-advice chatbot was evaluated more positively when it expressed sympathy or empathy rather than providing unemotional advice~\cite{Liu2018}. A systematic review also found that LLMs can demonstrate aspects of cognitive empathy, including recognizing emotional content and generating supportive responses~\cite{Sorin2024}. Ruben et al. further identify verbal features associated with perceived AI empathy, including validation, personalized active listening, nonjudgmental language, emotional reflection, and collaborative phrasing~\cite{Ruben2026}.

Given the success of empathic language in text-based and screen-based interactions, we examine verbal attunement during real-time spoken conversation with an embodied AI agent in VR. Unlike text chat, embodied VR interaction adds voice, conversational timing, spatial presence, facial expression, posture, and movement. We therefore test whether verbal attunement remains central to perceived empathy during face-to-face interaction.

\section{System Design: An Empathic Virtual Counselor}
In this section, we describe the implementation of the virtual counselor used in our study. We first outline the architecture supporting the conversational flow, followed by the implementation of verbal attunement, behavioral mimicry, and the embodied avatar. A system diagram summarizing the conversational and behavioral-mimicry pipelines, including their data flow and timing, is provided in the supplementary materials.

\subsection{Conversational Architecture and Interaction Flow}
The real-time conversation was handled through LiveKit\footnote{https://livekit.com}, which created a shared audio room between the VR application in Unity and a Python-based AI agent. A Node.js server generated the participant’s access token and assigned the agent to the same room. Once both were connected, the participant’s microphone audio was streamed to the agent. Silero VAD\footnote{https://github.com/snakers4/silero-vad} (voice activity detection) was used to identify when the participant was speaking, and Deepgram’s\footnote{https://deepgram.com/voice-ai-platform} Nova-2 model converted the speech into interim and final transcripts. Only final transcripts were treated as completed turns in conversation.

The counselor's task was to guide the participant through a short conversation structured around two study questions, described in more detail in the procedure section. After each completed response from the user, the system generated the counselor's next turn using the participant's transcript, the recent conversation history, the current study question, and the prompt instructions for the assigned experimental condition. This information was sent to GPT-4o mini, which generated a brief counselor response and one follow-up question. The structure of the session was controlled directly by the program, including the number of follow-up questions, transitions between main questions, the session time limit, and the closing sequence. This prevented the language model from independently changing the study procedure.

The generated response was then converted into speech using Deepgram’s Aura-2 Orion text-to-speech model and played through the embodied counselor in VR. After the agent finished speaking, the system returned to listening for the participant’s next response. This process repeated throughout the session, allowing the participant and virtual counselor to have a continuous spoken conversation in real time.
To contextualize system responsiveness, we conducted a post-hoc latency benchmark of the deployed STT--LLM--TTS pipeline. Across 12 logged turns, the mean latency from final ASR transcript receipt to counselor response initiation was 1.54 s. This delay is consistent with the architecture of pipeline-based voice agents, which accumulate latency across STT, LLM, TTS, and network stages, unlike real-time speech-to-speech systems with lower latency baselines~\cite{LiveKitPipelines,LiveKitRealtimeCascade}. In our system, responses were generated sequentially after ASR finalization rather than streamed across the full pipeline.

\begin{figure}[h]
    \centering
    \includegraphics[width=0.5\textwidth]{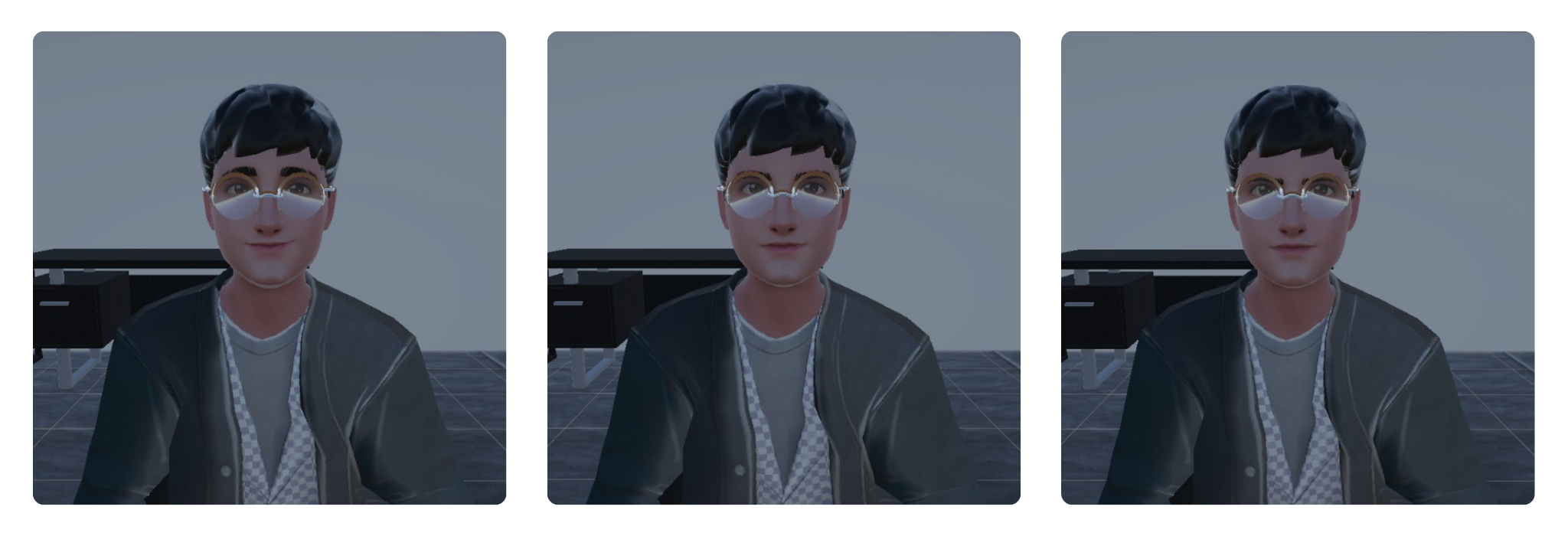}
    \caption{Counselor facial-expression states, shown from left to right: smiling, neutral, and brow lowering.}
    \label{facialMimicryImg}
\end{figure}

% \begin{figure}[h]
%     \centering
%     \includegraphics[width=0.5\textwidth]{HeadTilt.png}
%     \caption{The counselor in its default posture (left) and tilting its head to the left (right)}
%     \label{headTiltImg}
% \end{figure}

\subsection{Verbal Attunement}
The experiment included two versions of the conversational agent: a verbally attuned agent and a neutral agent. The agents used the same technical pipeline and followed the same overall interview structure. The primary difference between them was the language used in their prompts and how they responded to participant answers, asked follow-up questions, and transitioned between the main study questions.

The verbally attuned prompt was informed by Ruben et al.~\cite{Ruben2026}, who identified several verbal features associated with higher perceptions of empathy in LLM-based physician conversations. These included validation, reassurance, personalized active listening, nonjudgmental language, collaborative language, and attention to emotional or psychosocial information. Based on these findings, the agent was instructed to use brief validation, personalized active listening, nonjudgmental language, and collaborative wording and to offer light reassurance when appropriate. The prompt also instructed the agent to briefly reflect the participant's answer, recognize the underlying emotional meaning or personal value, and ask one relevant follow-up question. The agent was encouraged to refer to details provided by the participant rather than relying on generic responses. To verify that this manipulation produced a more verbally attuned agent, the study included a manipulation check assessing participants' perceptions of verbal attunement.

In contrast, the neutral agent was instructed to use professional, task-focused language without explicitly validating or interpreting the participant's emotions. It continued the conversation by acknowledging the participant's answer and asking a relevant follow-up question, but was not explicitly designed to use reassurance, emotional reflection, praise, and other explicitly empathic language as informed by Ruben et al. \cite{Ruben2026}. To support transparency and reproducibility, the complete system prompts for both verbal-attunement conditions and anonymized examples of responses generated during the study are provided in the supplementary materials.

\subsection{Behavioral Mimicry}
In the behavioral mimicry conditions, the counselor automatically mirrored a restricted set of participant behaviors: facial expressions, head tilts, and posture changes. These cues were selected because they were suitable for seated, face-to-face interaction and because prior work links facial and postural mimicry to empathy-related processes and social connection~\cite{Paiva2017, Hatfield1993, Niedenthal2007}, while head-movement mimicry can improve evaluations of virtual agents~\cite{Bailenson2005}. Detection remained active throughout the conversation, so mimicry could occur during both participant and counselor turns and was not gated by the active speaker. Facial-expression and posture changes had to remain stable for 30 consecutive frames before their onset or offset, helping filter brief or accidental movements, while head tilts used a one-second delay after crossing or returning below the detection threshold. This threshold was selected during iterative system testing as a practical balance between filtering brief or accidental facial and posture movements and maintaining responsive mimicry. In system testing, the observed end-to-end mimicry onset delays ranged from approximately 2 to 4 seconds, consistent with prior virtual-agent studies using delays between 1 and 4 seconds~\cite{Bailenson2005,Hale2016}.

During counselor speech, facial mimicry and lip synchronization operated concurrently but used separate mappings for emotional expressions and speech-related mouth movements. Although behavioral mimicry can involve continuous coordination across a broader range of facial, postural, and whole-body behaviors, this sensor-triggered implementation focused on cues that could naturally occur within the seated, face-to-face conversation used in our study.

\paragraph{\textbf{Facial Expression Mimicry}} The Meta Quest Pro captured participants’ facial movements in real time using Meta XR’s OVRFaceExpressions component. The system detected smiles by averaging the left and right lip-corner puller values and detected brow lowering by averaging the left and right brow-lowerer values. Each frame was classified as smiling when the smile average exceeded .05, as brow lowering when the smile threshold was not met and the brow-lowering average exceeded .20, or as neutral otherwise. We use “dominant facial expression” to refer to the resulting category, smile, brow lowering, or neutral, after the same classification remained stable for 30 consecutive frames. Once established, the counselor displayed the corresponding expression until another category met the same stability requirement. Detection remained active while participants spoke, with the 30-frame requirement helping reduce responses to brief speech-related mouth and cheek movements. Smiling and brow lowering were selected to capture positive and mildly negative affect, following prior work linking smiles to trustworthiness and cooperation and emotional-expression mimicry to emotion understanding and empathic resonance~\cite{Krumhuber2007,Niedenthal2007}.

\paragraph{\textbf{Head Tilt Mimicry}} The participant’s head orientation was tracked using the roll of the Meta Quest Pro headset. One second after initialization, the system recorded a neutral baseline and measured subsequent tilts relative to it each frame. A response was triggered when the tilt exceeded 10 degrees. The counselor matched the participant’s angle up to a maximum of 30 degrees, with leftward tilts mapped to the counselor’s right and vice versa to produce mirror matching. After a one-second delay, the counselor gradually moved toward the target angle and returned to neutral through the same process when the participant moved within the 10-degree threshold. The threshold and maximum angle were selected through iterative testing to distinguish intentional tilts while maintaining natural-looking movement.

\paragraph{\textbf{Posture Mimicry}}
At the beginning of each session, the system recorded the initial vertical position of the participant's headset as a reference for upright posture. During the conversation, the current headset height was compared with this starting position. When the participant's relative vertical position fell below a preset threshold, the system classified the participant as slouching. To avoid responding to brief or accidental posture changes, both the onset and offset of slouching had to remain stable for 30 consecutive frames before the counselor’s posture was updated. After a slouch was detected, the counselor's default upright idle animation was replaced with a slouched idle animation. When the participant returned above the threshold, the counselor returned to its standard idle posture.

\begin{figure}[h]
    \centering
    \includegraphics[width=0.5\textwidth]{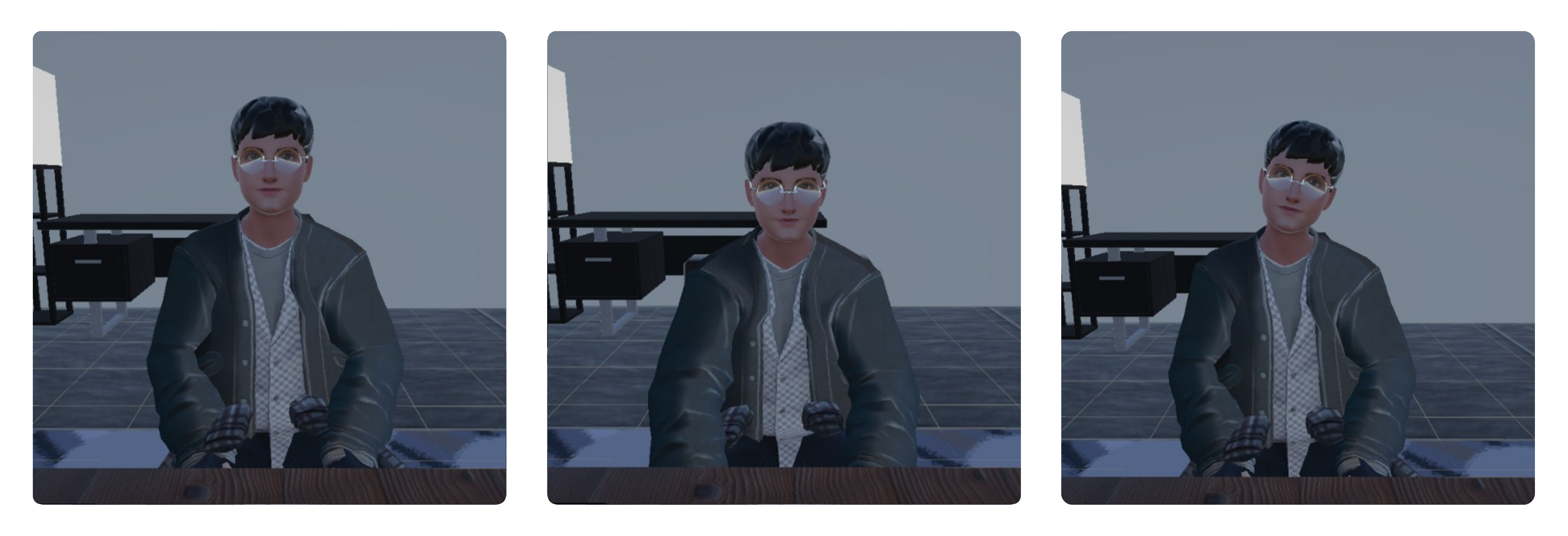}
    \caption{Counselor behaviors used for mimicry, shown from left to right: default upright posture, slouched posture, and leftward head tilt.}
    \label{postureImg}
\end{figure}

\subsection{Virtual Environment and Embodiment}
The virtual counselor was created using a Ready Player Me avatar with facial blendshapes that controlled its expressions and mouth movements. Lip synchronization was implemented using OVR LipSync, which mapped generated speech to the avatar’s mouth shapes. Facial mimicry used separate emotion mappings within the same component, so speech-related mouth movements and mirrored expressions were handled through distinct mappings. The complete blendshape mappings and intensity settings are provided in the supplementary materials.
% \begin{figure}[h]
%     \centering
%     \includegraphics[width=0.25\textwidth]{Counselor.png}
%     \caption{The Ready Player Me avatar used on the Sketchfab viewport}
%     \label{CounselorImg}
% \end{figure}

% \begin{figure}[h]
%     \centering
%     \includegraphics[width=0.5\textwidth]{Environment.png}
%     \caption{A screenshot of the virtual environment in Unity.}
%     \label{EnvironmentImg}
% \end{figure}

We also designed a simple counselor's office environment in Unity that matched the style of the avatar. The participant and virtual counselor were positioned across a table from one another to create a familiar face-to-face conversation setup. The counselor remained seated and used a neutral idle animation by default. In the behavioral mimicry conditions, this default animation and the avatar's facial or head movements could be modified in response to the participant's behavior.

\section{Study}

\begin{table*}[t]
\caption{Summary of 2 x 2 repeated-measures ANOVA tests for main and interaction effects of verbal attunement and behavioral mimicry. Bolded values denote statistically significant effects ($p < .05$) or marginal trends ($p < .10$).}
\label{tab:anova_results}
\centering
\begin{tabular}{|l|ccc|ccc|ccc|}
\hline
 & \multicolumn{3}{c|}{VA} 
 & \multicolumn{3}{c|}{BM} 
 & \multicolumn{3}{c|}{VA $\times$ BM} \\
\hline
 & $F$ & $p$ & $\eta_p^2$
 & $F$ & $p$ & $\eta_p^2$
 & $F$ & $p$ & $\eta_p^2$ \\
\hline
Empathy 
& \textbf{7.43} & \textbf{.013} & \textbf{.281}
& 1.28 & .273 & .063
& 0.69 & .418 & .035 \\

General Positivity 
& \textbf{3.47} & \textbf{.078} & \textbf{.154}
& 1.34 & .261 & .066
& 0.83 & .373 & .042 \\

Humanness 
& 1.26 & .275 & .062
& \textbf{4.31} & \textbf{.052} & \textbf{.185}
& 0.03 & .861 & .002 \\

Eeriness 
& 0.05 & .823 & .003
& 0.03 & .860 & .002
& 1.07 & .314 & .053 \\

\shortstack[l]{Verbal Attunement}
& \textbf{5.63} & \textbf{.028} & \textbf{.229}
& 2.58 & .125 & .119
& 0.33 & .571 & .017 \\
\hline
\end{tabular}
\end{table*}

\subsection{Participants}
We recruited 20 participants in the United States via listservs and flyers. All participants were pre-screened to ensure they fulfilled the eligibility requirements: being adults and fluent in English. 10 participants identified as male (50\%), eight as female (40\%), one as non-binary (5\%), and one preferred not to answer. Participant demographics can be found in the supplementary materials.

\subsection{Experimental Design}

We used a 2 x 2 within-subjects design with two experimental factors: verbal attunement and behavioral mimicry. Verbal attunement was either present or absent, and behavioral mimicry was either present or absent. This produced four counselor conditions: verbally attuned with mimicry (VA+BM), verbally attuned without mimicry (VA+NBM), neutral with mimicry (NVA+BM), and neutral without mimicry (NVA+NBM).

Each participant experienced all four conditions. To reduce order effects, we used a Latin square counterbalancing scheme. Participants were assigned to one of four counterbalancing groups, with five participants completing each condition order. Following Do et al.~\cite{Do2025}, who used affective avatars and Fast Friends-style prompts to support camaraderie-building conversation in remote settings, we used questions from Set II of the  Fast Friends procedure~\cite{Aron1997}. Before each participant's session, eight questions were randomly selected with two questions assigned to each condition. The questions in this set are designed to encourage moderately personal reflection and were used to keep the topic structure consistent across conditions while also giving the counselor opportunities to respond with empathy.

\subsection{Procedure}

Participants first completed a prescreening survey and demographic questionnaire. Eligible participants then scheduled an in-person study session with an experimenter. At the beginning of the session, the experimenter explained how to put on and adjust the VR headset and gave participants a brief overview of the task. Participants then completed four short conversations with the virtual counselor, one for each experimental condition. In each conversation, the counselor asked two of the selected Fast Friends questions and responded using the behavior assigned to that condition. At the end of each conversation, the counselor instructed the participant to remove the headset.

After each condition, participants completed a post-condition questionnaire about the counselor they had just interacted with. They then continued to the next condition until all four had been completed, following their assigned counterbalanced order.

\subsection{Measures}
Measures were collected after each counselor interaction as part of the post-condition survey. These survey measures captured participants' perceptions of the counselor they had just interacted with. In addition to these self-report measures, the system automatically logged mimicry exposure during each session. For transparency and reproducibility, the complete surveys, including all scale items and open-ended questions, are provided in the supplementary materials.

\subsubsection{Empathy and General Positivity}
Prior work has shown that empathy plays a critical role in long-lasting relationships ~\cite{Bickmore2005}. Accordingly, we measured perceived empathy using the Counselor Empathy Scale adapted from Guadagno et al. (2011)~\cite{Guadagno2011}, which asks participants to rate the agent across 12 interpersonal traits, including supportive, responsive, warm, sympathetic, and caring. In addition to empathy, we also wanted to examine whether verbal attunement and behavioral mimicry influenced participants' overall impression of the agent. To capture this broader response, we measured general positivity using the General Positivity Scale, also adapted from Guadagno et al.~\cite{Guadagno2011}.

\subsubsection{Humanness and Eeriness}
Because perceptions of avatar realism and uncanniness may influence users’ responses to embodied agents, we assessed participants’ perceptions of the counselor’s humanness and eeriness. Prior work has shown that highly humanlike virtual humans can evoke both increased perceptions of humanness and heightened feelings of eeriness, which are often associated with the uncanny valley phenomenon.

To measure these constructs, we adapted semantic differential items from the Uncanny Valley Index (UVI) developed by Ho and MacDorman~\cite{Ho2017} and used in recent virtual human research~\cite{Mal2024}. Participants rated the counselor avatar's humanness using three semantic differential items (Synthetic–Real, Mechanical Movement– Biological Movement, and Human-made–Humanlike) and its eeriness using two semantic differential items (Predictable–Eerie and Plain–Weird) using 7-point semantic differential scales. Responses were averaged separately to produce humanness and eeriness scores.

\subsubsection{Verbal Attunement Manipulation Check}
To assess whether the verbal attunement manipulation was successful, we developed a three-item manipulation check measuring participants' perceived verbal attunement of the AI agent. Items were administered after each condition on a 7-point Likert scale. The full item list is provided in the supplementary materials. The three items showed high internal consistency, Cronbach's $\alpha = .914$.

\subsubsection{Mimicry Exposure}
Because behavioral mimicry depended on participants producing detectable nonverbal behaviors, we tracked mimicry exposure during each session. We recorded the duration and frequency of mimicked behaviors and summed these durations to produce a total mimicry duration score for each session.

\subsubsection{Open-Ended Feedback}
The post-condition survey also included open-ended questions asking participants what they liked and disliked about the counselor. These responses were used to provide additional context for the quantitative results and to help identify patterns, concerns, or design issues that may not have been captured by the rating scales.

\section{Results}

\subsection{Quantitative Results}
\subsubsection{Data Analysis Approach}
\begin{figure*}[t]
\centering

% Row 1
\begin{subfigure}[t]{0.48\textwidth}
\centering
\begin{tikzpicture}
\begin{axis}[
    ybar,
    bar width=18pt,
    width=\linewidth,
    height=5cm,
    ymin=0,
    ymax=60,
    ylabel={Empathy},
    symbolic x coords={NVA,VA},
    xtick=data,
    tick label style={font=\small},
    label style={font=\small},
    enlarge x limits=0.35,
]
\addplot+[error bars/.cd, y dir=both, y explicit]
coordinates {
    (NVA,44.38) +- (0,2.54)
    (VA,51.30) +- (0,2.38)
};
\node[font=\scriptsize, blue, anchor=west] at (axis cs:NVA,47.2) {44.38};
\node[font=\scriptsize, blue, anchor=west] at (axis cs:VA,54.0) {51.30};
\end{axis}
\end{tikzpicture}
\caption{Empathy}
\label{fig:empathy_va_main}
\end{subfigure}
\hfill
\begin{subfigure}[t]{0.48\textwidth}
\centering
\begin{tikzpicture}
\begin{axis}[
    ybar,
    bar width=18pt,
    width=\linewidth,
    height=5cm,
    ymin=0,
    ymax=45,
    ylabel={General Positivity},
    symbolic x coords={NVA,VA},
    xtick=data,
    tick label style={font=\small},
    label style={font=\small},
    enlarge x limits=0.35,
]
\addplot+[error bars/.cd, y dir=both, y explicit]
coordinates {
    (NVA,36.20) +- (0,2.37)
    (VA,39.75) +- (0,2.11)
};
\node[font=\scriptsize, blue, anchor=west] at (axis cs:NVA, 38.2) {36.2};
\node[font=\scriptsize, blue, anchor=west] at (axis cs:VA,42.0) {39.75};
\end{axis}
\end{tikzpicture}
\caption{General Positivity}
\label{fig:positivity_va_main}
\end{subfigure}

% Row 2
\begin{subfigure}[t]{0.48\textwidth}
\centering
\begin{tikzpicture}
\begin{axis}[
    ybar,
    bar width=18pt,
    width=\linewidth,
    height=5cm,
    ymin=0,
    ymax=10,
    ylabel={Humanness},
    symbolic x coords={NBM,BM},
    xtick=data,
    tick label style={font=\small},
    label style={font=\small},
    enlarge x limits=0.35,
]
\addplot+[error bars/.cd, y dir=both, y explicit]
coordinates {
    (NBM,7.28) +- (0,0.72)
    (BM,8.63) +- (0,0.65)
};
\node[font=\scriptsize, blue, anchor=west] at (axis cs:NBM, 7.9) {7.28};
\node[font=\scriptsize, blue, anchor=west] at (axis cs:BM,9.1) {8.63};
\end{axis}
\end{tikzpicture}
\caption{Humanness}
\label{fig:humanness_bm_main}
\end{subfigure}
\hfill
\begin{subfigure}[t]{0.48\textwidth}
\centering
\begin{tikzpicture}
\begin{axis}[
    ybar,
    bar width=18pt,
    width=\linewidth,
    height=5cm,
    ymin=0,
    ymax=15,
    ylabel={Verbal Attunement},
    symbolic x coords={NVA,VA},
    xtick=data,
    tick label style={font=\small},
    label style={font=\small},
    enlarge x limits=0.35,
]
\addplot+[error bars/.cd, y dir=both, y explicit]
coordinates {
    (NVA,10.18) +- (0,0.81)
    (VA,12.00) +- (0,0.88)
};
\node[font=\scriptsize, blue, anchor=west] at (axis cs:NVA, 11) {10.18};
\node[font=\scriptsize, blue, anchor=west] at (axis cs:VA,13) {12.00};
\end{axis}
\end{tikzpicture}
\caption{Verbal Attunement}
\label{fig:verbal_attunement_va_main}
\end{subfigure}

\caption{Mean ratings for significant or marginal main effects. Empathy, general positivity, and verbal attunement are shown by verbal attunement condition, collapsed across behavioral mimicry. Humanness is shown by behavioral mimicry condition, collapsed across verbal attunement.}
\label{fig:main_effect_means}
\end{figure*}
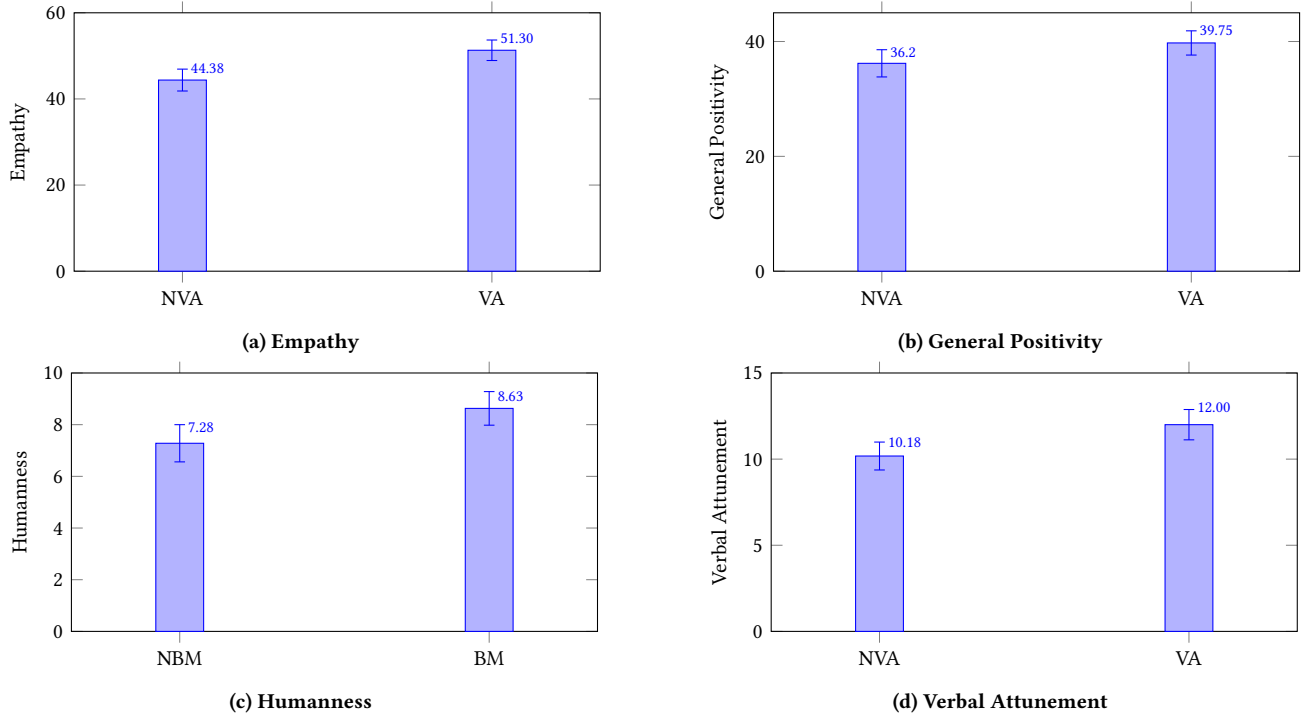

We conducted a series of 2 x 2 repeated-measures ANOVAs to examine the effects of verbal attunement and behavioral mimicry on participants' ratings of the counselor. Separate models were run for empathy, general positivity, humanness, eeriness, and the verbal attunement manipulation check. Pairwise comparisons of estimated marginal means were adjusted using the Bonferroni correction. Table~\ref{tab:anova_results} summarizes the main and interaction effects for all measures, while Table~\ref{tab:means} reports the means and standard deviations across the four experimental conditions.
Across these outcomes, verbal attunement and behavioral mimicry did not produce significant interaction effects, so the results below focus primarily on their main effects. Figure~\ref{fig:main_effect_means} visualizes the significant or marginal main effects, with each outcome grouped by the factor associated with that effect and collapsed across the other experimental factor. Across these panels, the VA conditions showed higher mean empathy, general positivity, and verbal-attunement ratings, while the BM conditions showed higher mean humanness.

\paragraph{\textbf{Empathy}}
We found a significant main effect of Verbal Attunement on Empathy, $F(1, 19) = 7.425, p = .013.$ Behavioral mimicry did not have a significant main effect on Empathy, $F(1, 19) = 1.277, p = .273.$

\paragraph{\textbf{General Positivity}}
We found a non-significant trend of Verbal Attunement on General Positivity, $F(1, 19) = 3.470, p = .078.$ Behavioral mimicry did not have a significant main effect on General Positivity, $F(1, 19) = 1.342, p = .261.$  

\paragraph{\textbf{Humanness}}
Verbal Attunement did not have a significant main effect on Humanness, $F(1, 19) = 1.264, p = .275.$ However, Behavioral mimicry showed a marginal effect on Humanness, $F(1, 19) = 4.314, p = .052.$ 

\paragraph{\textbf{Eeriness}}
Verbal Attunement did not have a significant main effect on Eeriness, $F(1, 19) = .051, p = .823.$ Similarly, Behavioral Mimicry also did not have a significant main effect on Eeriness, $F(1, 19) = .032, p = .86.$

\paragraph{\textbf{Verbal Attunement Manipulation Check}}
We found a significant main effect of Verbal Attunement on the Verbal Attunement Manipulation Check, $F(1, 19) = 5.634, p = .028.$ Behavioral mimicry did not have a significant main effect on the Verbal Attunement Manipulation Check, $F(1, 19) = 2.576, p = .125.$ 

\subsubsection{Exploratory Analysis of Mimicry Exposure by Gender}
Because the counselor could only mimic behaviors that participants actually displayed, mimicry exposure varied across sessions. Some participants naturally smiled, tilted their head, or slouched more than others, while others performed no mimicable behaviors. For this reason, we examined whether total mirroring duration was associated with counselor ratings. Across all participants, mirroring duration was positively but not significantly correlated with empathy, $r(38) = .241, p = .135$, general positivity, $r(38) = .269, p = .093$, and humanness, $r(38) = .294, p = .065$.

Prior work has reported small average gender differences in the recognition of affective and nonverbal social cues~\cite{Hall2013, Hall2025, Thompson2014}. Motivated by this literature, we conducted exploratory gender-stratified Pearson correlations to examine whether associations between mirroring exposure and counselor ratings showed different patterns by participant gender. Among female participants, mirroring duration was significantly positively correlated with empathy, $r(14) = .530, p = .035$, general positivity, $r(14) = .555, p = .025$, and humanness, $r(14) = .620, p = .010$. Among male participants, mirroring duration was not significantly correlated with empathy, $r(18) = .108, p = .649$, general positivity, $r(18) = .171, p = .472$, or humanness, $r(18) = .198, p = .403$. Because these analyses were exploratory and subgroup sample sizes were small, these patterns should be interpreted cautiously.
\begin{table}[t]
\centering
\caption{Means and standard deviations for all measurements. Values are $M$ ($SD$).}
\label{tab:means}
\resizebox{\columnwidth}{!}{%
\begin{tabular}{lcccc}
\hline
Measure & NVA+NBM & NVA+BM & VA+NBM & VA+BM \\ 
\hline
Empathy   & 42.00 (15.38) & 46.75 (13.48) & 50.90 (10.57) & 51.70 (13.74) \\
Positivity & 34.30 (13.14) & 38.10 (12.49) & 39.40 (10.08) & 40.10 (10.28) \\
Humanness & 6.95 (3.75)   & 8.40 (3.30)   & 7.60 (3.40)   & 8.85 (3.45)   \\
Eeriness  & 7.55 (3.32)   & 7.00 (2.70)   & 6.95 (2.70)   & 7.20 (2.19)   \\
VA Check  & 9.40 (4.49)   & 10.95 (4.38)  & 11.60 (4.42)  & 12.40 (4.26)  \\ 
\hline
\end{tabular}%
}
\end{table}
\subsection{Open-Ended Observations}
We conducted a qualitative analysis of participants' responses to the open-ended post-condition survey questions. Following Braun and Clarke's six-phase framework, we used an inductive thematic approach to identify recurring patterns in participants' experiences with the counselors. Responses were first reviewed for recurring ideas, then grouped into preliminary themes based on their relevance to the study, such as perceived emotional support, conversational flow, embodied behavior, and moments of discomfort or distraction. These themes were refined iteratively as the analysis progressed. The first author coded the responses, while the remaining authors provided guidance in interpreting and refining the themes, supporting rigor in line with McDonald's recommendations for qualitative analysis in HCI research ~\cite{McDonald2019}.

\subsubsection{\textbf{Verbal Attunement Reduced Perceived Coldness and Increased Attunement}}
Participants' responses showed more coded comments describing emotional distance or coldness in the NVA conditions than in the VA conditions. Across the two NVA conditions, we identified 10 comments describing coldness or emotional distance, compared with 4 such comments across the two VA conditions. After experiencing an NVA condition, P19 stated, "\textit{It was much more interrogative and less supportively reactive. The main goal seemed to be to solicit information to its benefit rather than my own.}" In contrast, comments describing responsiveness, emotional attunement, or involvement were more frequent in the VA conditions (14 comments) than in the NVA conditions (5 comments). After experiencing a VA session, P13 noted, "\textit{[The counselor] told me it appreciated what I shared and used specific words I used. This made me feel seen and encouraged to be more open to the counselor.}" Similarly, P10 wrote, "\textit{The counselor would repeat back what I shared and rephrase it to ask the next question, something that makes the counselor seem more involved.}"

\subsubsection{\textbf{Interaction Breakdowns Limited the Perceived Benefits of Attunement}}

Across conditions, participants evaluated the counselor not only by what it said, but also by how well the interaction unfolded. Interruptions, abrupt topic changes, long pauses, and responses that seemed to miss or cut off participants' answers often reduced the counselor's perceived responsiveness, even in verbally attuned conditions. For example, P11 wrote, "\textit{It felt like they were rushing me a little bit with my responses}" Similarly, P3 stated, "\textit{it didn't seem responsive. sometimes didnt catch everything i said or started talking while i paused to think in the middle of a response.}" 

\subsubsection{\textbf{Embodied Behavior Helped When It Signaled Attentiveness, But Hurt When It Appeared Mechanical}}
Participants' comments suggest that embodied behavior was most helpful when it made the counselor appear attentive and available, but less effective when it drew attention to the agent's artificiality. In mimicry conditions, some participants explicitly described the counselor's posture and body language in positive ways. For example, P17 wrote, "\textit{I liked that the counselor seemed ready to listen and his posture was inviting.}" In another mimicry condition, the same participant noted, "\textit{Something I liked about this counselor was body language and the attentiveness that he seemed to have with his body language.}" However, embodied behavior could also become distracting when it appeared unnatural or mechanical. P12 stated that "\textit{The initial head movement was a bit eerie, looks a bit contrived,}" while P3 noted that "\textit{the mouth movements looked unnatural.}" Other participants described the counselor's body language as rigid or mechanically strange. These comments suggest that mimicry and embodied movement did not uniformly improve the interaction. Instead, embodied behavior seemed beneficial when it supported the counselor's perceived attentiveness, but harmful when it appeared uncanny, poorly synchronized, or animated in a way that drew attention to itself.

\section{Discussion}

\subsection{Verbal Attunement Was the Primary Driver of Perceived Empathy}
The strongest effect in our study was the effect of verbal attunement on perceived empathy. Although the counselor was visually embodied and could mimic participants' nonverbal behavior in some conditions, participants' empathy ratings were most consistently shaped by how the counselor responded to what they said. This suggests that, when users disclose personal experiences to an embodied AI counselor, perceived empathy depends strongly on whether the agent seems to understand and respond to the user's specific disclosure.

This finding extends prior work on empathic language by showing its importance in a real-time embodied AI setting where both verbal and nonverbal cues were present. Verbal attunement appeared to provide the clearest signal that the counselor was listening and responding with care. In this sense, perceived empathy depended less on whether the counselor produced synchronized nonverbal behavior, and more on whether its responses felt specific, relevant, and emotionally responsive.

Participants' open-ended responses reflected this pattern. In verbally attuned conditions, participants more often described the counselor as responsive, involved, or emotionally aware. For example, P13 wrote, “\textit{The counselor reflected back to me feelings I said, which was surprising and felt good.}” Similarly, P17 noted, “\textit{I liked this counselor more. it seems like he was actually listening to me and the responses were better. The responses were actually related to what I was saying before.}" These comments suggest that participants interpreted empathy through moments where the counselor seemed to understand and build on their specific response, rather than through general friendliness or behavioral responsiveness alone.

\subsection{Behavioral Mimicry Supported the Counselor's Humanlike Presence}
Behavioral mimicry showed a marginal effect on perceived humanness ($p=0.052$). In this study, humanness mattered because the counselor was designed for an emotionally reflective interaction rather than a purely task-based exchange. Participants were asked to disclose personal experiences, so the agent's role depended not only on providing responses, but also on feeling socially present enough to participate in a face-to-face conversational setting. Behavioral mimicry may have supported this perception by making the counselor feel less static and more responsive as an embodied presence.

However, humanness should not be treated as an unconditional design goal. For VR-based AI agents in socially oriented roles, the goal is not necessarily to make the agent indistinguishable from a human, but to make its humanlike cues coherent enough to support comfort, attention, and social engagement. 

Participants' open-ended comments reflected this distinction. For example, one participant in a mimicry condition noted that the counselor's posture made it seem “\textit{ready to listen},” “\textit{inviting}," and “\textit{attentive}”. Rather than showing that mimicry directly increased disclosure, these comments suggest that behavioral mimicry may have helped establish a more disclosure-supportive interaction by making the counselor appear attentive, available, and socially present. In contrast, a participant in a non-mimicry condition noted that although the counselor's body language seemed attentive, its lack of movement made the interaction feel "\textit{off-putting}." Together, these comments suggest that contingent embodied responses, such as mimicry, may help socially oriented VR agents feel more present when they make the agent appear responsive to the user, rather than merely animated.

\subsection{Mimicry Exposure Showed Exploratory Gender-Stratified Patterns}
Contrary to our expectations, the restricted set of categorical mimicry cues implemented in this study did not have a significant main effect on empathy or general positivity. This finding should not be interpreted as evidence that behavioral mimicry broadly is ineffective. Rather, it characterizes the effects of the specific facial-expression, head-tilt, and posture cues evaluated within this VR counseling context. Accordingly, this is consistent with prior work showing that the effects of virtual mimicry are not always reliable in controlled VR settings~\cite{Hale2016}. Hale and Hamilton argue that mimicry is difficult to study in isolation because real-world social interaction includes many overlapping cues, such as gaze, facial expression, timing, turn-taking, and emotional responsiveness. In our study, mimicry exposure was positively, but not significantly, correlated with empathy, general positivity, and humanness ratings across the full sample.

When we examined mimicry exposure by gender, a more specific pattern appeared. Among participants who identified as female, mirroring duration was significantly positively correlated with empathy, general positivity, and humanness. Because the subgroup sample sizes were small and gender was not a primary factor in the study design, these results should be interpreted as exploratory patterns that may warrant examination in a larger, appropriately powered study. Still, the pattern is consistent with prior work showing that women often show greater sensitivity to affective and nonverbal social cues, including small advantages in recognizing nonverbal emotional displays and context-dependent advantages in empathic accuracy~\cite{Thompson2014,Hall2013,Hall2025}. In this study, mimicry may have been more noticeable or meaningful for participants who were more sensitive to subtle nonverbal cues of responsiveness.

\subsection{Turn-Taking and Conversational Flow Shaped Perceived Responsiveness}
Because this study involved real-time spoken interaction, participants evaluated the counselor not only by the content of its responses, but also by the timing and flow of the conversation. Responsiveness is a core part of perceived empathy, and in spoken interaction it depends on more than generating an attuned response. The system must also determine when the user has finished speaking, when a pause is meaningful, and when it is appropriate to begin the next turn.

This creates a design tension for real-time AI agents. Responses need to arrive quickly enough for the conversation to feel natural, but not so quickly that a reflective pause is treated as the end of the user's turn. Several participants described moments where the counselor responded before they had finished speaking. For example, P3 wrote, "\textit{it didn't seem responsive. Sometimes it didnt catch everything I said or started talking while I paused to think in the middle of a response.}" Similarly, P7 noted, "\textit{The responsiveness felt weird, I got cut-off mid sentence multiple times. I don't think it fully registered my responses before responding.}"

These comments suggest that turn-taking errors can undermine perceived responsiveness even when the agent's language is designed to be empathic. For emotionally reflective VR agents, pauses are not simply silence to be minimized; they may indicate that the user is thinking, choosing how much to disclose, or preparing a more personal response. Designers should therefore treat turn-taking and end-of-utterance detection as part of the empathy design, rather than just a technical layer supporting the conversation.

\subsection{Limitations and Future Work}
This study points to several opportunities for future work. First, although the verbal-attunement and non-attuned conditions were designed to differ primarily in their degree of emotional acknowledgment and reflection, they were not explicitly matched for response length or speaking duration. Because the verbal-attunement prompt encouraged more reflective responses, differences in length or elaborateness may have contributed alongside verbal attunement to participants’ empathy ratings. Future studies could further disentangle these factors by using length-matched responses or accounting for word count and speaking duration in the analysis. Second, because the system supported real-time spoken interaction, participants' perceptions of responsiveness were shaped not only by what the counselor said, but also by the timing of conversational turns. In emotionally reflective conversations, pauses may indicate that a participant is thinking or deciding how much to disclose, rather than that they have finished speaking. Future systems should therefore continue improving turn-taking and end-of-utterance detection as part of empathic interaction design. Third, behavioral mimicry exposure was uneven across participants because mimicry depended on whether participants naturally produced detectable behaviors such as smiling, head tilting, or posture changes. This means that condition-level comparisons may not fully capture how much mimicry participants actually experienced, and future work should continue to examine duration, frequency, and type of mimicry exposure. Fourth, the gender-stratified findings were exploratory and based on small subgroup samples, so larger studies are needed to determine whether user gender reliably moderates responses to behavioral mimicry. Fifth, our behavioral-mimicry manipulation used a restricted set of categorical facial-expression, head-tilt, and posture cues. Although these cues were selected based on prior literature and their suitability for seated interaction, they do not capture the full range or continuous dynamics of facial and whole-body mimicry. Future work should examine richer and more context-sensitive forms of nonverbal mirroring.  Finally, this study focused on an embodied AI counselor in an emotionally reflective setting. Future work should examine whether these findings extend to other VR-based AI agents, such as tutors, collaborators, interviewers, or companions, and should investigate the role of race and identity in perceived empathy, including whether participants perceive greater empathy from agents that share their racial or cultural identity.

\section{Conclusion}
Our findings show how verbal and nonverbal cues contribute differently to the design of empathic AI agents in VR. Verbal attunement had the clearest effect on perceived empathy, suggesting that users primarily judged the counselor's empathy through responses that expressed support, reflected emotional meaning, and addressed their specific disclosures. Behavioral mimicry played a different role. Rather than directly increasing empathy ratings, the implemented mimicry cues showed a marginal effect on perceived humanness, while greater mimicry exposure was positively associated with empathy, general positivity, and humanness. These results suggest that even a limited set of contingent embodied cues may contribute to the counselor’s humanlike presence, while highlighting the need to examine richer forms of behavioral mimicry. For designers, the findings also highlight the importance of real-time conversational flow: empathic responses must be delivered at the right moment, without cutting off users' reflective pauses or disrupting their sense of being heard. Overall, this work suggests that multimodal synchrony remains a promising design direction for VR-based AI agents, but its effects depend on how verbal attunement, embodied behavior, and conversational timing work together.

\bibliographystyle{ACM-Reference-Format}
\bibliography{sample-base}

@article{Nam2025,
	title = {{AR} {Fitness} {Dog}: {Effects} of a {User}-{Mimicking} {Interactive} {Virtual} {Pet} on {User} {Experience} and {Social} {Presence} in {Physical} {Exercise}},
	volume = {31},
	issn = {1941-0506},
	shorttitle = {{AR} {Fitness} {Dog}},
	url = {https://ieeexplore.ieee.org/abstract/document/10919006},
	doi = {10.1109/TVCG.2025.3549858},
	number = {5},
	urldate = {2026-06-04},
	journal = {IEEE Transactions on Visualization and Computer Graphics},
	author = {Nam, Hyeongil and Lee, Kisub and Park, Jong-II and Kim, Kangsoo},
	month = may,
	year = {2025},
	pages = {2817--2827},
}

@article{liu_screens_2025,
	title = {From screens to scenes: {A} survey of embodied {AI} in healthcare},
	volume = {119},
	issn = {1566-2535},
	shorttitle = {From screens to scenes},
	url = {https://www.sciencedirect.com/science/article/pii/S156625352500106X},
	doi = {10.1016/j.inffus.2025.103033},
	urldate = {2026-06-04},
	journal = {Information Fusion},
	author = {Liu, Yihao and Cao, Xu and Chen, Tingting and Jiang, Yankai and You, Junjie and Wu, Minghua and Wang, Xiaosong and Feng, Mengling and Jin, Yaochu and Chen, Jintai},
	month = jul,
	year = {2025},
	pages = {103033},
}

@article{mir_embodied_2026,
	title = {Embodied {Artificial} {Intelligence} in {Healthcare}: {A} {Systematic} {Review} of {Robotic} {Perception}, {Decision}-{Making}, and {Clinical} {Impact}},
	volume = {14},
	copyright = {http://creativecommons.org/licenses/by/3.0/},
	issn = {2227-9032},
	shorttitle = {Embodied {Artificial} {Intelligence} in {Healthcare}},
	url = {https://www.mdpi.com/2227-9032/14/5/572},
	doi = {10.3390/healthcare14050572},
	language = {en},
	number = {5},
	urldate = {2026-06-04},
	journal = {Healthcare},
	publisher = {Multidisciplinary Digital Publishing Institute},
	author = {Mir, Bilal Ahmad and Nishwa, Dur E. and Lee, Seung Won},
	month = jan,
	year = {2026},
	pages = {572},
}

@inproceedings{rings_ivas_2025,
	address = {New York, NY, USA},
	series = {{MuC} '25},
	title = {{IVAs}' {Future} in {Therapy}: {Investigating} the {Impact} of {Intelligent} {Virtual} {Agents} in {VR} on {Psychological} {Self}-{Help} {Exercises} and {User} {Satisfaction}},
	isbn = {979-8-4007-1582-2},
	shorttitle = {{IVAs}' {Future} in {Therapy}},
	url = {https://dl.acm.org/doi/10.1145/3743049.3743063},
	doi = {10.1145/3743049.3743063},
	urldate = {2026-06-04},
	booktitle = {Proceedings of the {Mensch} und {Computer} 2025},
	publisher = {Association for Computing Machinery},
	author = {Rings, Sebastian and Kruse, Lucie and Rudschies, Catharina and Rolvien, Lara and Schauenburg, Gesche and Gallinat, Jürgen and Schneider, Ingrid and Steinicke, Frank},
	month = aug,
	year = {2025},
	pages = {193--209},
}

@inproceedings{yu_embodied_2026,
	address = {New York, NY, USA},
	series = {{IUI} '26},
	title = {Embodied {Digital} {Therapists} with {LLM} {Personalization} for {Aphasia} {Rehabilitation}: {Characterizing} {Human}-{AI} {Collaboration} {Boundaries}},
	isbn = {979-8-4007-1984-4},
	shorttitle = {Embodied {Digital} {Therapists} with {LLM} {Personalization} for {Aphasia} {Rehabilitation}},
	url = {https://dl.acm.org/doi/10.1145/3742413.3789090},
	doi = {10.1145/3742413.3789090},
	urldate = {2026-06-05},
	booktitle = {Proceedings of the 31st {International} {Conference} on {Intelligent} {User} {Interfaces}},
	publisher = {Association for Computing Machinery},
	author = {Yu, Mengting and Zhu, Lifeng and Chen, Wenli and Liu, Jin and Liu, Zhaoyi and Song, Aiguo},
	month = mar,
	year = {2026},
	pages = {941--954},
}

@article{kim_revisiting_2018,
	title = {Revisiting {Trends} in {Augmented} {Reality} {Research}: {A} {Review} of the 2nd {Decade} of {ISMAR} (2008–2017)},
	volume = {24},
	issn = {1941-0506},
	shorttitle = {Revisiting {Trends} in {Augmented} {Reality} {Research}},
	url = {https://ieeexplore.ieee.org/abstract/document/8456568},
	doi = {10.1109/TVCG.2018.2868591},
	number = {11},
	urldate = {2026-06-05},
	journal = {IEEE Transactions on Visualization and Computer Graphics},
	author = {Kim, Kangsoo and Billinghurst, Mark and Bruder, Gerd and Duh, Henry Been-Lirn and Welch, Gregory F.},
	month = nov,
	year = {2018},
	pages = {2947--2962},
}

@inproceedings{kruse_would_2023,
	title = {Would {You} {Go} to a {Virtual} {Doctor}? {A} {Systematic} {Literature} {Review} on {User} {Preferences} for {Embodied} {Virtual} {Agents} in {Healthcare}},
	issn = {2473-0726},
	shorttitle = {Would {You} {Go} to a {Virtual} {Doctor}?},
	url = {https://ieeexplore.ieee.org/abstract/document/10316479},
	doi = {10.1109/ISMAR59233.2023.00082},
	urldate = {2026-06-05},
	booktitle = {2023 {IEEE} {International} {Symposium} on {Mixed} and {Augmented} {Reality} ({ISMAR})},
	author = {Kruse, Lucie and Hertel, Julia and Mostajeran, Fariba and Schmidt, Susanne and Steinicke, Frank},
	month = oct,
	year = {2023},
	note = {ISSN: 2473-0726},
	pages = {672--682},
}

@article{weidner_systematic_2023,
	title = {A {Systematic} {Review} on the {Visualization} of {Avatars} and {Agents} in {AR} \& {VR} displayed using {Head}-{Mounted} {Displays}},
	volume = {29},
	issn = {1941-0506},
	url = {https://ieeexplore.ieee.org/abstract/document/10049669},
	doi = {10.1109/TVCG.2023.3247072},
	number = {5},
	urldate = {2026-06-05},
	journal = {IEEE Transactions on Visualization and Computer Graphics},
	author = {Weidner, Florian and Boettcher, Gerd and Arboleda, Stephanie Arevalo and Diao, Chenyao and Sinani, Luljeta and Kunert, Christian and Gerhardt, Christoph and Broll, Wolfgang and Raake, Alexander},
	month = may,
	year = {2023},
	pages = {2596--2606},
}

@article{winkler_avatar_2025,
	title = {Avatar and virtual agent-assisted telecare for patients in their homes: {A} scoping review},
	volume = {31},
	issn = {1357-633X},
	shorttitle = {Avatar and virtual agent-assisted telecare for patients in their homes},
	url = {https://doi.org/10.1177/1357633X231174484},
	doi = {10.1177/1357633X231174484},
	language = {EN},
	number = {2},
	urldate = {2026-06-05},
	journal = {Journal of Telemedicine and Telecare},
	publisher = {SAGE Publications},
	author = {Winkler, Anna and Kutschar, Patrick and Pitzer, Stefan and van der Zee-Neuen, Antje and Kerner, Susanne and Osterbrink, Jürgen and Krutter, Simon},
	month = mar,
	year = {2025},
	pages = {207--221},
}

@inproceedings{kim_reducing_2020,
	title = {Reducing {Task} {Load} with an {Embodied} {Intelligent} {Virtual} {Assistant} for {Improved} {Performance} in {Collaborative} {Decision} {Making}},
	issn = {2642-5254},
	url = {https://ieeexplore.ieee.org/abstract/document/9089596},
	doi = {10.1109/VR46266.2020.00074},
	urldate = {2026-06-05},
	booktitle = {2020 {IEEE} {Conference} on {Virtual} {Reality} and {3D} {User} {Interfaces} ({VR})},
	author = {Kim, Kangsoo and de Melo, Celso M. and Norouzi, Nahal and Bruder, Gerd and Welch, Gregory F.},
	month = mar,
	year = {2020},
	note = {ISSN: 2642-5254},
	pages = {529--538},
}

@inproceedings{earle-randell_how_2025,
	address = {Cham},
	title = {How {Virtual} {Agents} {Can} {Shape} {Human}-{Human} {Collaboration}: {A} {Systematic} {Review}},
	isbn = {978-3-031-98420-4},
	shorttitle = {How {Virtual} {Agents} {Can} {Shape} {Human}-{Human} {Collaboration}},
	doi = {10.1007/978-3-031-98420-4_33},
	language = {en},
	booktitle = {Artificial {Intelligence} in {Education}},
	publisher = {Springer Nature Switzerland},
	author = {Earle-Randell, Toni V. and Zhang, Shan and Schroeder, Noah and Boyer, Kristy E. and Dorley, Emmanuel},
	editor = {Cristea, Alexandra I. and Walker, Erin and Lu, Yu and Santos, Olga C. and Isotani, Seiji},
	year = {2025},
	pages = {468--486},
}

@inproceedings{do_effects_2022,
	address = {New York, NY, USA},
	series = {{ICMI} '22},
	title = {The {Effects} of an {Embodied} {Pedagogical} {Agent}’s {Synthetic} {Speech} {Accent} on {Learning} {Outcomes}},
	isbn = {978-1-4503-9390-4},
	url = {https://dl.acm.org/doi/10.1145/3536221.3556587},
	doi = {10.1145/3536221.3556587},
	urldate = {2026-06-05},
	booktitle = {Proceedings of the 2022 {International} {Conference} on {Multimodal} {Interaction}},
	publisher = {Association for Computing Machinery},
	author = {Do, Tiffany D. and Akter, Mamtaj and Choudhary, Zubin and Azevedo, Roger and McMahan, Ryan P.},
	month = nov,
	year = {2022},
	pages = {198--206},
}

@article{volante_effects_2016,
	title = {Effects of {Virtual} {Human} {Appearance} {Fidelity} on {Emotion} {Contagion} in {Affective} {Inter}-{Personal} {Simulations}},
	volume = {22},
	issn = {1941-0506},
	url = {https://ieeexplore.ieee.org/abstract/document/7383334},
	doi = {10.1109/TVCG.2016.2518158},
	number = {4},
	urldate = {2026-06-05},
	journal = {IEEE Transactions on Visualization and Computer Graphics},
	author = {Volante, Matias and Babu, Sabarish V. and Chaturvedi, Himanshu and Newsome, Nathan and Ebrahimi, Elham and Roy, Tania and Daily, Shaundra B. and Fasolino, Tracy},
	month = apr,
	year = {2016},
	pages = {1326--1335},
}

@inproceedings{kim_dominance_2026,
	address = {New York, NY, USA},
	series = {{CHI} '26},
	title = {The {Dominance} {Effect}: {How} {Verbal} and {Nonverbal} {Cues} of {Virtual} {Agents} {Influence} {Decision}-{Making} in {VR}},
	isbn = {979-8-4007-2278-3},
	shorttitle = {The {Dominance} {Effect}},
	url = {https://dl.acm.org/doi/10.1145/3772318.3791539},
	doi = {10.1145/3772318.3791539},
	urldate = {2026-06-05},
	booktitle = {Proceedings of the 2026 {CHI} {Conference} on {Human} {Factors} in {Computing} {Systems}},
	publisher = {Association for Computing Machinery},
	author = {Kim, Taeyeon and Nam, Hyeongil and Jung, Sunghun and A Fouad, Ahmad and Kim, Kangsoo and Lee, Myungho},
	month = apr,
	year = {2026},
	pages = {1--14},
}

@inproceedings{sonlu_effects_2025,
	title = {Effects of {Embodiment} and {Personality} in {LLM}-{Based} {Conversational} {Agents}},
	issn = {2642-5254},
	url = {https://ieeexplore.ieee.org/abstract/document/10937459},
	doi = {10.1109/VR59515.2025.00094},
	urldate = {2026-06-05},
	booktitle = {2025 {IEEE} {Conference} {Virtual} {Reality} and {3D} {User} {Interfaces} ({VR})},
	author = {Sonlu, Sinan and Bendiksen, Bennie and Durupinar, Funda and Güdükbay, Uğur},
	month = mar,
	year = {2025},
	note = {ISSN: 2642-5254},
	pages = {718--728},
}

@article{Ho2017,
	title = {Measuring the {Uncanny} {Valley} {Effect}: {Refinements} to {Indices} for {Perceived} {Humanness}, {Attractiveness}, and {Eeriness}},
	volume = {9},
	issn = {1875-4791, 1875-4805},
	shorttitle = {Measuring the {Uncanny} {Valley} {Effect}},
	url = {http://link.springer.com/10.1007/s12369-016-0380-9},
	doi = {10.1007/s12369-016-0380-9},
	language = {en},
	number = {1},
	urldate = {2026-06-18},
	journal = {International Journal of Social Robotics},
	author = {Ho, Chin-Chang and MacDorman, Karl F.},
	month = jan,
	year = {2017},
	pages = {129--139},
}

@inproceedings{Mal2024,
	address = {Honolulu HI USA},
	title = {From {2D}-{Screens} to {VR}: {Exploring} the {Effect} of {Immersion} on the {Plausibility} of {Virtual} {Humans}},
	isbn = {979-8-4007-0331-7},
	shorttitle = {From {2D}-{Screens} to {VR}},
	url = {https://dl.acm.org/doi/10.1145/3613905.3650773},
	doi = {10.1145/3613905.3650773},
	language = {en},
	urldate = {2026-06-18},
	booktitle = {Extended {Abstracts} of the {CHI} {Conference} on {Human} {Factors} in {Computing} {Systems}},
	publisher = {ACM},
	author = {Mal, David and Wolf, Erik and Döllinger, Nina and Botsch, Mario and Wienrich, Carolin and Latoschik, Marc Erich},
	month = may,
	year = {2024},
	pages = {1--8},
}

@article{Guadagno2011,
	title = {Social evaluations of embodied agents and avatars},
	volume = {27},
	issn = {07475632},
	url = {https://linkinghub.elsevier.com/retrieve/pii/S0747563211001555},
	doi = {10.1016/j.chb.2011.07.017},
	language = {en},
	number = {6},
	urldate = {2025-11-18},
	journal = {Computers in Human Behavior},
	author = {Guadagno, Rosanna E. and Swinth, Kimberly R. and Blascovich, Jim},
	month = nov,
	year = {2011},
	pages = {2380--2385},
}

@article{McDonald2019,
	title = {Reliability and {Inter}-rater {Reliability} in {Qualitative} {Research}: {Norms} and {Guidelines} for {CSCW} and {HCI} {Practice}},
	volume = {3},
	issn = {2573-0142},
	shorttitle = {Reliability and {Inter}-rater {Reliability} in {Qualitative} {Research}},
	url = {https://dl.acm.org/doi/10.1145/3359174},
	doi = {10.1145/3359174},
	language = {en},
	number = {CSCW},
	urldate = {2023-07-25},
	journal = {Proceedings of the ACM on Human-Computer Interaction},
	author = {McDonald, Nora and Schoenebeck, Sarita and Forte, Andrea},
	month = nov,
	year = {2019},
	pages = {1--23},
}

@article{Hale2016,
	title = {Testing the relationship between mimicry, trust and rapport in virtual reality conversations},
	volume = {6},
	issn = {2045-2322},
	url = {https://www.nature.com/articles/srep35295},
	doi = {10.1038/srep35295},
	language = {en},
	number = {1},
	urldate = {2026-06-21},
	journal = {Scientific Reports},
	author = {Hale, Joanna and Hamilton, Antonia F. De C.},
	month = oct,
	year = {2016},
	pages = {35295},
}

@article{Thompson2014,
	title = {Sex differences in the ability to recognise non-verbal displays of emotion: {A} meta-analysis},
	volume = {28},
	issn = {0269-9931, 1464-0600},
	shorttitle = {Sex differences in the ability to recognise non-verbal displays of emotion},
	url = {http://www.tandfonline.com/doi/abs/10.1080/02699931.2013.875889},
	doi = {10.1080/02699931.2013.875889},
	language = {en},
	number = {7},
	urldate = {2026-06-21},
	journal = {Cognition and Emotion},
	author = {Thompson, Ashley E. and Voyer, Daniel},
	month = oct,
	year = {2014},
	pages = {1164--1195},
}

@incollection{Hall2013,
	title = {21 {Gender} differences in nonverbal communication},
	isbn = {978-3-11-023814-3 978-3-11-023815-0},
	url = {https://www.degruyterbrill.com/document/doi/10.1515/9783110238150.639/html},
	doi = {10.1515/9783110238150.639},
	language = {en},
	urldate = {2026-06-21},
	booktitle = {Nonverbal {Communication}},
	publisher = {DE GRUYTER},
	author = {Hall, Judith A. and Gunnery, Sarah D.},
	editor = {Hall, Judith A. and Knapp, Mark L.},
	month = jan,
	year = {2013},
	pages = {639--670},
}

@article{Hall2025,
	title = {Gender and {Accuracy} in {Decoding} {Affect} {Cues}: {A} {Meta}-{Analysis}},
	volume = {13},
	issn = {2079-3200},
	shorttitle = {Gender and {Accuracy} in {Decoding} {Affect} {Cues}},
	url = {https://www.mdpi.com/2079-3200/13/3/38},
	doi = {10.3390/jintelligence13030038},
	language = {en},
	number = {3},
	urldate = {2026-06-21},
	journal = {Journal of Intelligence},
	author = {Hall, Judith A. and Gunnery, Sarah D. and Schlegel, Katja},
	month = mar,
	year = {2025},
	pages = {38},
}

@article{Yang2025,
	title = {Embodied {Conversational} {Agents} in {Extended} {Reality}: {A} {Systematic} {Review}},
	volume = {13},
	copyright = {https://creativecommons.org/licenses/by-nc-nd/4.0/},
	issn = {2169-3536},
	shorttitle = {Embodied {Conversational} {Agents} in {Extended} {Reality}},
	url = {https://ieeexplore.ieee.org/document/10985757/},
	doi = {10.1109/ACCESS.2025.3566698},
	language = {en},
	urldate = {2026-06-25},
	journal = {IEEE Access},
	author = {Yang, Fu-Chia and Acevedo, Pedro and Guo, Siqi and Choi, Minsoo and Mousas, Christos},
	year = {2025},
	pages = {79805--79824},
}

@misc{Fung2025,
	title = {Embodied {AI} {Agents}: {Modeling} the {World}},
	shorttitle = {Embodied {AI} {Agents}},
	url = {http://arxiv.org/abs/2506.22355},
	doi = {10.48550/arXiv.2506.22355},
	language = {en},
	urldate = {2025-11-13},
	publisher = {arXiv},
	author = {Fung, Pascale and Bachrach, Yoram and Celikyilmaz, Asli and Chaudhuri, Kamalika and Chen, Delong and Chung, Willy and Dupoux, Emmanuel and Gong, Hongyu and Jégou, Hervé and Lazaric, Alessandro and Majumdar, Arjun and Madotto, Andrea and Meier, Franziska and Metze, Florian and Morency, Louis-Philippe and Moutakanni, Théo and Pino, Juan and Terver, Basile and Tighe, Joseph and Tomasello, Paden and Malik, Jitendra},
	month = jul,
	year = {2025},
	note = {arXiv:2506.22355 [cs]},
}

@article{Paiva2017,
	title = {Empathy in {Virtual} {Agents} and {Robots}: {A} {Survey}},
	volume = {7},
	issn = {2160-6455, 2160-6463},
	shorttitle = {Empathy in {Virtual} {Agents} and {Robots}},
	url = {https://dl.acm.org/doi/10.1145/2912150},
	doi = {10.1145/2912150},
	language = {en},
	number = {3},
	urldate = {2025-11-11},
	journal = {ACM Transactions on Interactive Intelligent Systems},
	author = {Paiva, Ana and Leite, Iolanda and Boukricha, Hana and Wachsmuth, Ipke},
	month = sep,
	year = {2017},
	pages = {1--40},
}

@article{Bailenson2005,
	title = {Digital {Chameleons}: {Automatic} {Assimilation} of {Nonverbal} {Gestures} in {Immersive} {Virtual} {Environments}},
	volume = {16},
	copyright = {https://journals.sagepub.com/page/policies/text-and-data-mining-license},
	issn = {0956-7976, 1467-9280},
	shorttitle = {Digital {Chameleons}},
	url = {https://journals.sagepub.com/doi/10.1111/j.1467-9280.2005.01619.x},
	doi = {10.1111/j.1467-9280.2005.01619.x},
	language = {en},
	number = {10},
	urldate = {2025-11-11},
	journal = {Psychological Science},
	author = {Bailenson, Jeremy N. and Yee, Nick},
	month = oct,
	year = {2005},
	pages = {814--819},
}

@article{Ruben2026,
	title = {What is {Artificial} {Intelligence} ({AI}) “{Empathy}”? {A} {Study} {Comparing} {ChatGPT} and {Physician} {Responses} on an {Online} {Forum}},
	volume = {41},
	issn = {0884-8734, 1525-1497},
	shorttitle = {What is {Artificial} {Intelligence} ({AI}) “{Empathy}”?},
	url = {https://link.springer.com/10.1007/s11606-025-10068-w},
	doi = {10.1007/s11606-025-10068-w},
	language = {en},
	number = {5},
	urldate = {2026-06-17},
	journal = {Journal of General Internal Medicine},
	author = {Ruben, Mollie A. and Blanch-Hartigan, Danielle and Hall, Judith A.},
	month = apr,
	year = {2026},
	pages = {1304--1311},
}

@article{Bickmore2005,
	title = {Establishing and maintaining long-term human-computer relationships},
	volume = {12},
	issn = {1073-0516, 1557-7325},
	url = {https://dl.acm.org/doi/10.1145/1067860.1067867},
	doi = {10.1145/1067860.1067867},
	language = {en},
	number = {2},
	urldate = {2025-11-13},
	journal = {ACM Transactions on Computer-Human Interaction},
	author = {Bickmore, Timothy W. and Picard, Rosalind W.},
	month = jun,
	year = {2005},
	pages = {293--327}
}

@article{Lakin2003,
author={Lakin,Jessica L. and Jefferis,Valerie E. and Cheng,Clara M. and Chartrand,Tanya L.},
year={2003},
month={Fall},
title={The chameleon effect as social glue: Evidence for the evolutionary significance of nonconscious mimicry},
journal={Journal of Nonverbal Behavior},
volume={27},
number={3},
pages={145-162},
isbn={01915886},
language={English},
}

@article{deBorst2015,
	title = {Is it the real deal? {Perception} of virtual characters versus humans: an affective cognitive neuroscience perspective},
	volume = {6},
	issn = {1664-1078},
	shorttitle = {Is it the real deal?},
	url = {http://www.frontiersin.org/Cognitive_Science/10.3389/fpsyg.2015.00576/abstract},
	doi = {10.3389/fpsyg.2015.00576},
	language = {en},
	urldate = {2026-06-23},
	journal = {Frontiers in Psychology},
	author = {De Borst, Aline W. and De Gelder, Beatrice},
	month = may,
	year = {2015},
}

@article{Liu2018,
	title = {Should {Machines} {Express} {Sympathy} and {Empathy}? {Experiments} with a {Health} {Advice} {Chatbot}},
	volume = {21},
	url = {https://journals.sagepub.com/doi/abs/10.1089/cyber.2018.0110},
	doi = {10.1089/cyber.2018.0110},
	number = {10},
	journal = {Cyberpsychology, Behavior, and Social Networking},
	author = {Liu, Bingjie and Sundar, S. Shyam},
	year = {2018},
    pages = {625--636},
}

@article{Sorin2024,
	title = {Large {Language} {Models} and {Empathy}: {Systematic} {Review}},
	volume = {26},
	issn = {1438-8871},
	shorttitle = {Large {Language} {Models} and {Empathy}},
	url = {https://www.jmir.org/2024/1/e52597},
	doi = {10.2196/52597},
	language = {en},
	urldate = {2026-06-23},
	journal = {Journal of Medical Internet Research},
	author = {Sorin, Vera and Brin, Dana and Barash, Yiftach and Konen, Eli and Charney, Alexander and Nadkarni, Girish and Klang, Eyal},
	month = dec,
	year = {2024},
	pages = {e52597},
}

@article{Ayers2023,
	title = {Comparing {Physician} and {Artificial} {Intelligence} {Chatbot} {Responses} to {Patient} {Questions} {Posted} to a {Public} {Social} {Media} {Forum}},
	volume = {183},
	issn = {2168-6106},
	url = {https://jamanetwork.com/journals/jamainternalmedicine/fullarticle/2804309},
	doi = {10.1001/jamainternmed.2023.1838},
	language = {en},
	number = {6},
	urldate = {2026-06-23},
	journal = {JAMA Internal Medicine},
	author = {Ayers, John W. and Poliak, Adam and Dredze, Mark and Leas, Eric C. and Zhu, Zechariah and Kelley, Jessica B. and Faix, Dennis J. and Goodman, Aaron M. and Longhurst, Christopher A. and Hogarth, Michael and Smith, Davey M.},
	month = jun,
	year = {2023},
	pages = {589},
}

@inproceedings{Do2025,
	address = {Yokohama Japan},
	title = {Exploring {AI}-{Driven} {Affective} {Avatars} for {Autistic} {Adults} and {Adults} with {Social} {Anxiety} in {Virtual} {Meetings}},
	isbn = {979-8-4007-1395-8},
	url = {https://dl.acm.org/doi/10.1145/3706599.3719885},
	doi = {10.1145/3706599.3719885},
	language = {en},
	urldate = {2026-06-24},
	booktitle = {Proceedings of the {Extended} {Abstracts} of the {CHI} {Conference} on {Human} {Factors} in {Computing} {Systems}},
	publisher = {ACM},
	author = {Do, Tiffany D. and Mott, Martez E and Tang, John and Junuzovic, Sasa and Paradiso, Ann and Cutrell, Edward},
	month = apr,
	year = {2025},
	pages = {1--9},
}

@article{Aron1997,
author = {Arthur Aron and Edward Melinat and Elaine N. Aron and Robert Darrin Vallone and Renee J. Bator},
title ={The Experimental Generation of Interpersonal Closeness: A Procedure and Some Preliminary Findings},
journal = {Personality and Social Psychology Bulletin},
volume = {23},
number = {4},
pages = {363-377},
year = {1997},
doi = {10.1177/0146167297234003},
URL = { 
        https://doi.org/10.1177/0146167297234003
},
eprint = { 
        https://doi.org/10.1177/0146167297234003
}
}

@misc{LiveKitPipelines,
  author = {{LiveKit}},
  title = {Pipeline types},
  year = {2026},
  howpublished = {\url{https://docs.livekit.io/agents/models/pipelines/}},
  note = {Accessed: 2026-06-24}
}

@misc{LiveKitRealtimeCascade,
  author = {{LiveKit}},
  title = {Pipeline vs. Realtime: Which is the Better Voice Agent Architecture?},
  year = {2026},
  howpublished = {\url{https://livekit.com/blog/realtime-vs-cascade}},
  note = {Accessed: 2026-06-24}
}

@article{Hatfield1993,
author = {Elaine Hatfield and John T. Cacioppo and Richard L. Rapson},
title ={Emotional Contagion},
journal = {Current Directions in Psychological Science},
volume = {2},
number = {3},
pages = {96-100},
year = {1993},
doi = {10.1111/1467-8721.ep10770953},
URL = { 
        https://doi.org/10.1111/1467-8721.ep10770953
},
eprint = { 
        https://doi.org/10.1111/1467-8721.ep10770953
}

}

@article{Elliott2018,
	title = {Therapist empathy and client outcome: {An} updated meta-analysis.},
	volume = {55},
	issn = {1939-1536, 0033-3204},
	shorttitle = {Therapist empathy and client outcome},
	url = {https://doi.apa.org/doi/10.1037/pst0000175},
	doi = {10.1037/pst0000175},
	language = {en},
	number = {4},
	urldate = {2026-06-24},
	journal = {Psychotherapy},
	author = {Elliott, Robert and Bohart, Arthur C. and Watson, Jeanne C. and Murphy, David},
	month = dec,
	year = {2018},
	pages = {399--410},
}

@article{Niedenthal2007,
	title = {Embodying {Emotion}},
	volume = {316},
	issn = {0036-8075, 1095-9203},
	url = {https://www.science.org/doi/10.1126/science.1136930},
	doi = {10.1126/science.1136930},
	language = {en},
	number = {5827},
	urldate = {2026-06-24},
	journal = {Science},
	author = {Niedenthal, Paula M.},
	month = may,
	year = {2007},
	pages = {1002--1005},
}

@article{Krumhuber2007,
	title = {Facial dynamics as indicators of trustworthiness and cooperative behavior.},
	volume = {7},
	issn = {1931-1516, 1528-3542},
	url = {https://doi.apa.org/doi/10.1037/1528-3542.7.4.730},
	doi = {10.1037/1528-3542.7.4.730},
	language = {en},
	number = {4},
	urldate = {2026-04-06},
	journal = {Emotion},
	author = {Krumhuber, Eva and Manstead, Antony S. R. and Cosker, Darren and Marshall, Dave and Rosin, Paul L. and Kappas, Arvid},
	year = {2007},
	pages = {730--735},
}

\appendix
\section*{Supplemental Materials}

All supplemental materials are available on Zenodo at \url{https://zenodo.org/records/22307230}, released under a CC BY 4.0 license. They include the screening and post-experimental surveys; a participant demographics table; the verbal-attunement manipulation-check items; technical details of the facial-expression mimicry implementation; complete system prompts and example responses for the verbally attuned and neutral agents; a system diagram; and a supplementary video.

% \section{Verbal Attunement Check Questions}
% \label{VAQuestions}
% \begin{itemize}
% \item The counselor seemed to understand what I was saying.
% \item The counselor’s responses felt emotionally appropriate.
% \item The counselor responded in a way that fit the tone of what I shared. 
% \end{itemize}

% \section{Participant Demographics}
% \begin{table}[H]
% \caption{Participant demographics.}
% \label{tab:participant_demographics}
% \begin{tabular}{lll|lll}
% \hline
% \textbf{P\#} & \textbf{Age} & \textbf{Gender} & \textbf{P\#} & \textbf{Age} & \textbf{Gender} \\ \hline
% p0           & 30           & F               & p1           & 22           & F               \\
% p2           & 26           & M               & p3           & 25           & M               \\
% p4           & 21           & M               & p5           & 21           & M               \\
% p6           & 24           & M               & p7           & 23           & F               \\
% p8           & 21           & NB              & p9           & 24           & F               \\
% p10          & 20           & F               & p11          & 21           & M               \\
% p12          & 26           & F               & p13          & 31           & PNTA            \\
% p14          & 21           & M               & p15          & 23           & M               \\
% p16          & 23           & M               & p17          & 26           & F               \\
% p18          & 25           & F               & p19          & 28           & M              
% \end{tabular}
% \end{table}

\end{document}